\documentclass[aps,preprint,lengthcheck,prb,showpacs,noshowkeys,superscriptaddress]{revtex4-1}%
\usepackage{amssymb}
\usepackage{amsfonts}
\usepackage{amsmath}
\usepackage{graphicx}
\usepackage{natbib}%
\providecommand{\U}[1]{\protect\rule{.1in}{.1in}}

\begin{document}
\preprint{HEP/123-qed}
\title[Short title for running header]{Evolution of the Phase Diagram of LaFeP$_{1-x}$As$_{x}$O$_{1-y}$%
F$_{y}$ ($y$ = 0 -- 0.1)}
\author{K. T. Lai}
\affiliation{Department of Physics, Osaka University, Osaka 560-0043, Japan}
\author{A. Takemori}
\affiliation{Department of Physics, Osaka University, Osaka 560-0043, Japan}
\author{S. Miyasaka}\email[]{e-mail address: miyasaka@phys.sci.osaka-u.ac.jp}
\affiliation{Department of Physics, Osaka University, Osaka 560-0043, Japan}
\author{F. Engetsu}
\affiliation{Graduate School of Engineering Science, Osaka University, Osaka 560-8531, Japan}
\author{H. Mukuda}
\affiliation{Graduate School of Engineering Science, Osaka University, Osaka 560-8531, Japan}
\author{S. Tajima}
\affiliation{Department of Physics, Osaka University, Osaka 560-0043, Japan}

\keywords{iron-based superconductors, doping dependence, transport properties}
\pacs{74.70.Xa, 74.25.F-, 74.62.Dh}

\begin{abstract}
P/As-substitution effects on the transport properties of polycrystalline
LaFeP$_{1-x}$As$_{x}$O$_{1-y}$F$_{y}$ with $x$ = 0 -- 1.0 and $y$ = 0 -- 0.1 have
been studied. In the F-free samples ($y$ = 0), a new superconducting (SC) dome with a maximum $T_{c}$ of 12 K
is observed around $x$ = 0 -- 0.3. This is separated from another SC dome with $T_{c}$ $\sim$10 K at $x$ = 0.6 -- 0.8 by an antiferromagnetic region ($x$ = 0.3 -- 0.6), giving a two-dome feature in the $T_{c}-x$ phase diagram. As $y$ increases, the two SC domes merge together, changing to a double peak structure at $y$ = 0.05 and a single dome at $y$ = 0.1. This proves the presence of two different Fermi surface states in this system. 

\end{abstract}
\volumeyear{year}
\volumenumber{number}
\issuenumber{number}
\eid{identifier}
\date[Date: ]{\today}
\received[Received text]{date}

\revised[Revised text]{date}

\accepted[Accepted text]{date}

\published[Published text]{date}

\startpage{101}
\endpage{102}
\maketitle

\section{Introduction}
Since the discovery of iron-based superconductors, one of the focuses is to find a unified picture of the electronic state among various kinds of iron-based superconductors. It has
been revealed that the suppression of spin density wave (SDW) in their parent
compounds is important to induce high-$T_{c}$ superconductivity
(SC) in most of the cases. For example, hole doped (Ba,K)Fe$_{2}$As$_{2}$
\cite{Rotter2008}, electron doped Ba(Fe,Co)$_{2}$As$_{2}$ \cite{Chu2009} and
isovalent-element doped BaFe$_{2}$(As,P)$_{2}$ \cite{Kasahara2010} share a similar
phase diagram which shows the transition from SDW to SC state upon doping. This similarity suggests that the spin
fluctuation (SF) near a quantum critical point (QCP) is a possible candidate for
the pairing force of Cooper pairs. In LaFeAsO, however, the phase diagram is
much more complicated. For example, in LaFeAs(O,H)
\cite{Iimura2012,Fujiwara2013}, as H content increases after the suppression of SDW, two SC regions have been observed in the phase diagram, as well as the reappearance of an
antiferromagnetic (AFM) phase in the overdoped region. The origin of these
interesting behaviors has been argued in relation to SF or orbital
fluctuations \cite{Fujiwara2013,Iimura2013,Yamakawa2013,Suzuki2013,Onari2014}%
.\nolinebreak

On the other hand, LaFePO is
known as a superconductor with $T_{c}$ $\sim$5 K without any trace of
SDW \cite{Kamihara2006,Hamlin2008}. Introducing electrons by F substitution cannot change
$T_{c}$ significantly \cite{Suzuki2009}. The striking difference between the
properties of the two parent compounds LaFeAsO and LaFePO raises a question about
its origin and the relation to SC. Experimental observations
\cite{Coldea2008,Sugawara2008,Liu2010,Li2009,Lu2008} and theoretical
calculations
\cite{Lebe`gue2007,Singh2008,Kuroki2009,Vildosola2008,Thomale2011} have shown
that the Fermi surface (FS) topologies of LaFeAsO and LaFePO are different, mainly in one of
the hole pockets. In particular, the 2-dimensional FS exists in LaFeAsO around X-point, while the 3-dimensional one around Z-point exists in LaFePO. The difference in the resulting FS nesting may be related to the
different SC behaviors. 

P/As substitution in LaFe(As,P)O provides a platform for studying the relationship among SC, AFM and corresponding FS nesting. A study of LaFe(As,P)O in P doping level from 0 to 60\% has been reported previously
\cite{Wang2009}. After the suppression of SDW around LaFeAsO, a SC dome with a maximum
critical temperature $T_{c}$ $\sim$10 K appears around 30\% P doping. Recently, a further study from NMR technique has indicated the AFM ordering around 50\% P doping with $T_{N}$ $\sim$15 K \cite{Kitagawa2014}. Such complex behaviors suggest the importance of the role of AFM correlation to induce SC in LaFe(As,P)O.

In our previous work, we have also studied P/As substitution effect in the same system but for 10\% F-doping, namely, LaFeP$_{1-x}$As$_{x}%
$O$_{0.9}$F$_{0.1}$ \cite{Saijo2010,Miyasaka2011,Miyasaka2013}. A maximum $T_{c}$
$\sim$28 K has been observed at $x$ = 0.6 together with temperature ($T$)-linearly
dependent resistivity and strong $T$ dependent Hall coefficient. It
suggests SF  is strong around $x$ = 0.6, which has been confirmed in the NMR study \cite{Mukuda2014}. Moreover, it has been revealed that this anomaly at $x$ = 0.6 are commonly observed not only in La-1111 system but also in Nd-1111 and Pr-1111 system which are different in lattice size. This implies that the important electronic change, presumably a band crossover, is driven by P/As substitution. 

As described above, LaFePO and LaFeAsO systems have different FS states because of the existence of $d_{Z^{2}}$ or $d_{X^{2}-Y^{2}}$ band near Fermi level. The P/As substitution causes the exchange of the energy level of $d_{Z^{2}}$/$d_{X^{2}-Y^{2}}$ band. Resultantly, our previous results show the anomalous behaviors, suggesting that $R$FeP$_{1-x}$As$_{x}%
$O$_{0.9}$F$_{0.1}$ ($R$ = La, Pr and Nd) has the FeP-type FS below $x \sim$ 0.6, while the samples of $x$ = 1.0 have FeAs-type FS. However, the evidence of our scenario (two FS states and the band crossover around $x$ = 0.6) was weak. 

In this study, we have extended our previous work to lower F-content ($y$ = 0 and 0.05). The transport properties for LaFeP$_{1-x}$As$_{x}$O$_{1-y}$F$_{y}$ with $y$ = 0 and 0.05 have been mainly studied, and we have tried to clarify the phase diagram, which provides further evidence for the presence of two electronic states originating from the two different FS topologies and its crossover.  

\section{Experimental Methods and sample characterization}
Polycrystalline LaFeP$_{1-x}$As$_{x}$O$_{1-y}$F$_{y}$ with nominal $x$ = 0
-- 1 and $y$ = 0, 0.05 were synthesized by a solid state reaction method.
The precursors LaAs/LaP, Fe$_{2}$O$_{3}$, Fe and LaF$_{3}$ were
mixed and pressed into pellets.
The pellets were sealed in evacuated quartz tubes and heated at 1100 $%
\operatorname{{}^{\circ}{\rm C}}%
$ for 40 h. All the above processes except heating were performed in a glove box with pure Ar environment. 

The crystal structure was characterized by powder X-ray
diffraction with the source of Cu \textit{K}$_{\alpha}$ radiation at room
temperature. Magnetic susceptibility measurements were performed in a Quantum Design MPMS
with the applied field of 10 Oe. The electrical resistivity was measured by a standard
four-probe method. Hall effect measurements were
performed in the magnetic field up to 7 T. $R_{H}$ was obtained from Hall resistivity which showed linear dependence on the magnetic applied field. The $^{31}$P-NMR spectra in the AFM phase were measured by sweeping the magnetic field to determine the transition temperature $T_{N}$ \cite{Mukuda2}.

\begin{figure}[h]%
\centering
\includegraphics[scale=0.3]%
{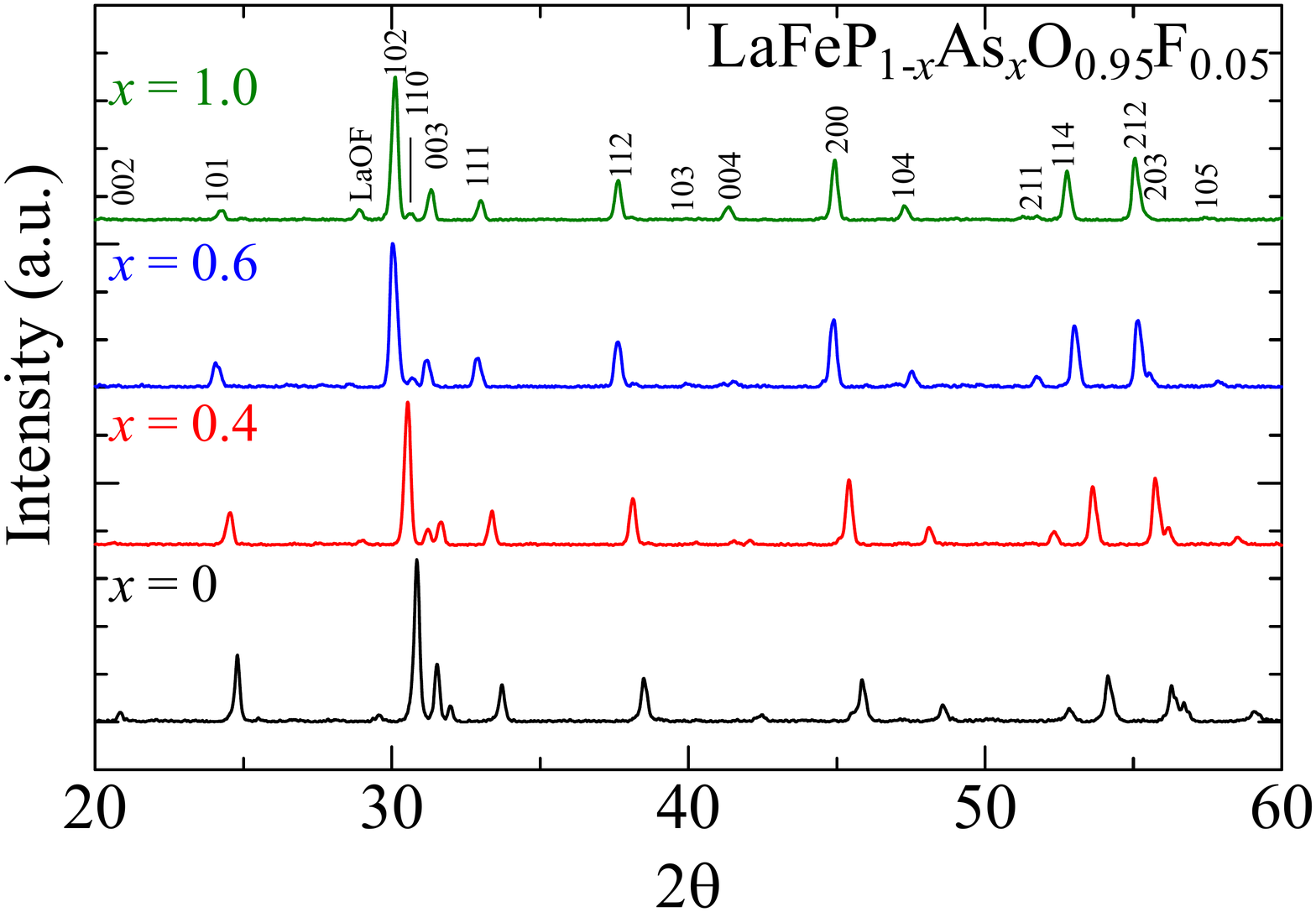}%
\caption{(Color online) The X-ray diffraction patterns of
LaFeP$_{1-x}$As$_{x}$O$_{0.95}$F$_{0.05}$ ($x$ = 0, 0.4, 0.6 and 1.0). Almost all the peaks are able to be indexed assuming the $P$4/$nmm$ tetragonal symmetry.}%
\label{s1}%
\end{figure}

\begin{figure}[h]%
\centering
\includegraphics[scale=0.3]%
{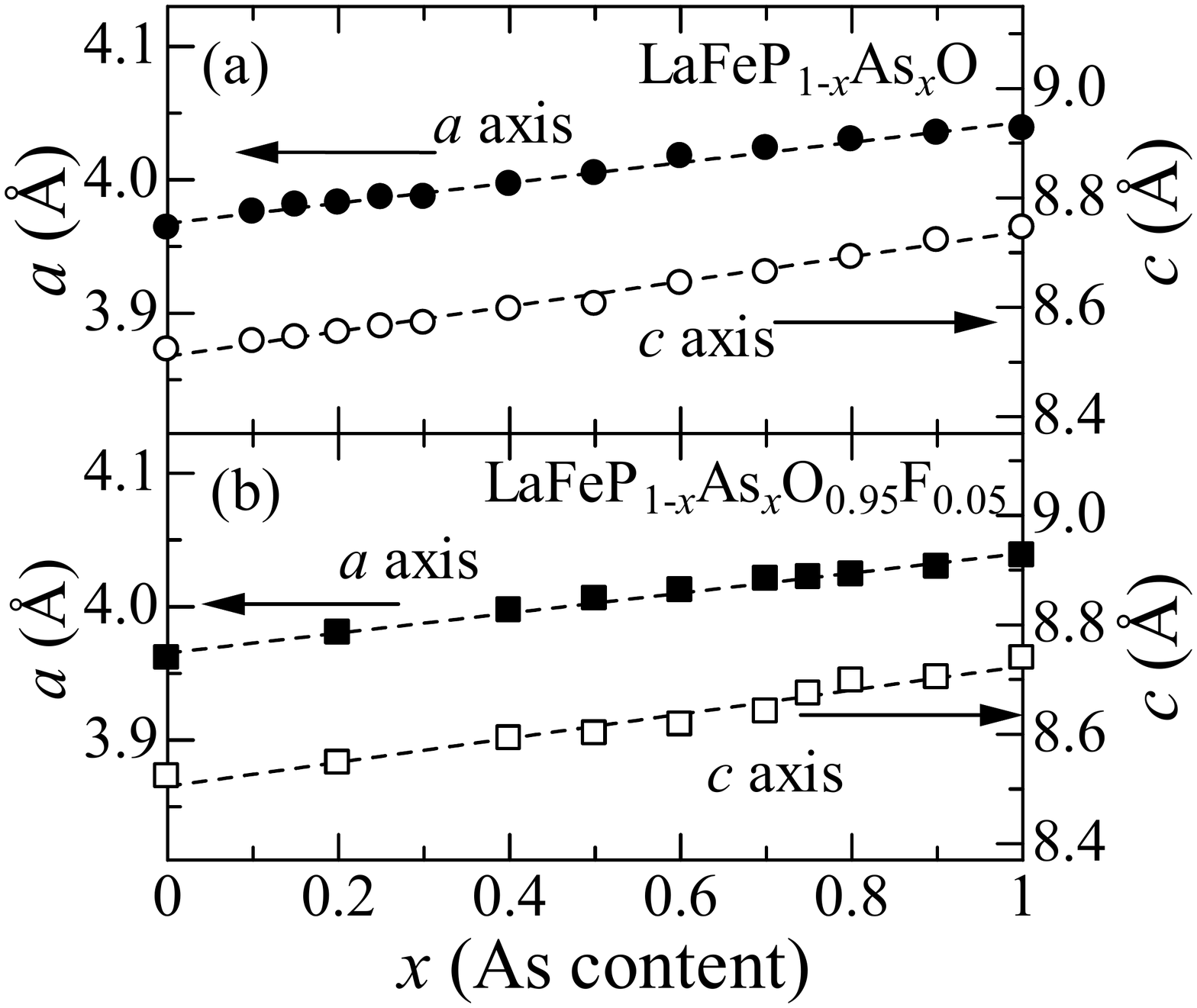}%
\caption{The $x$ dependence of the lattice constants $a$ and $c$ of
(a) LaFeP$_{1-x}$As$_{x}$O and (b) LaFeP$_{1-x}$As$_{x}$O$_{0.95}$F$_{0.05}$.}%
\label{s2}%
\end{figure}

The typical powder X-ray diffraction patterns for $y$ = 0.05 are shown in Fig.~\ref{s1}. Almost all the Bragg peaks observed in the diffraction patterns are able to be assigned within the tetragonal $P4/nmm$ symmetry. Note that a minor impurity peak of LaOF is found in some samples. The
corresponding lattice constants $a$ and $c$ for the samples of $y$ = 0 and 0.05 are calculated by the
least-square fitting of the Bragg peaks and the data is plotted in Fig.
\ref{s2}. Both samples for $y$ = 0 and 0.05 show a linear increase in $a$ and $c$ with increasing As content
$x$. This linear change of lattice constants indicates
that the As/P solution compounds are successfully prepared.

In order to determine the actual As and F concentrations, we performed the energy dispersive X-ray spectroscopy (EDX) measurements. The result of EDX indicates that the actual As concentration is the same as the nominal one ($x$) in all the samples. On the other hand, we are not able to estimate the actual F concentration by EDX measurements, because there are peaks for La and Fe near the peak for F in the EDX spectrum. With increasing the nominal F concentration $y$, however, the lattice constants $a$ and $c$ continuously decrease. (Please see Fig. \ref{s2} and Ref. [27].) In the previous report, we have roughly estimated the actual F concentration of $\sim$0.03 -- 0.04 in the samples with the nominal F concentration $y $= 0.1 \cite{Miyasaka2013}. Assuming that the lattice constants depend linearly on the actual F concentration, the actual F concentration is about 0.01 in the samples with nominal F concentration of $y$ = 0.05. In short, we can determine the actual As concentration and not the actual F one by EDX measurements. For convenience, we use the nominal F concentration ($y$ = 0 and 0.05) in this paper.

\section{Results and Discussions}

\begin{figure}[h]%
\centering
\includegraphics[scale=0.09]%
{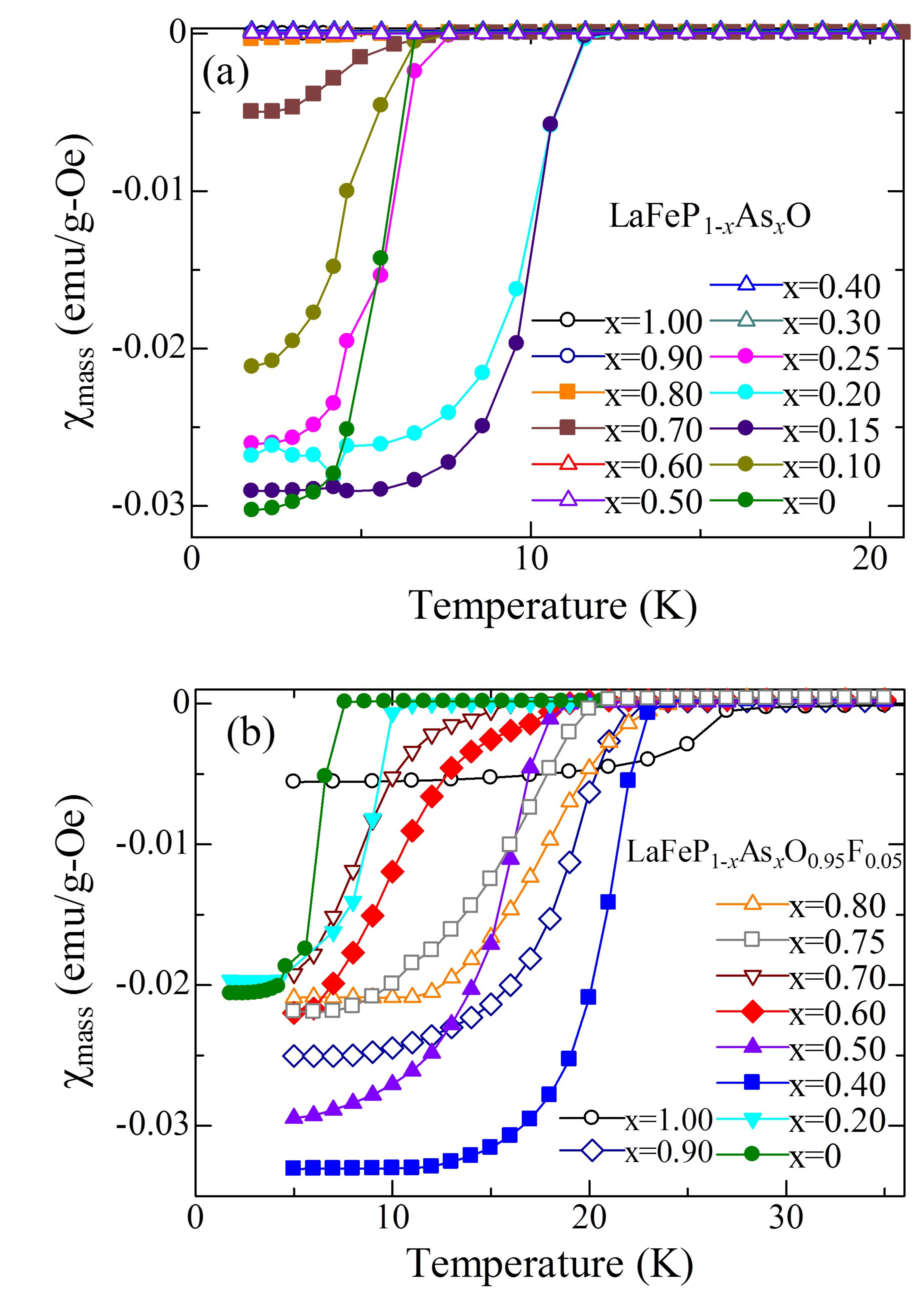}%
\caption{(Color online) The temperature dependence of magnetic susceptibility of (a) LaFeP$_{1-x}$As$_{x}$O
and (b) LaFeP$_{1-x}$As$_{x}$O$_{0.95}$F$_{0.05}$.}%
\label{s3}%
\end{figure}

Figure \ref{s3} shows the $T$-dependence of magnetic susceptibility for
the samples of $y$ = 0 and 0.05. The SC transition can be
observed in all the samples of $y$ = 0.05, and the samples of $y$ = 0 with $x$ = 0
-- 0.25, 0.7 and 0.8. Note that the superconducting volume fraction of
the samples of $x$ = 0.7 and 0.8 for $y$ = 0 is much smaller than the other SC samples.

\begin{figure}[ptb]
\centering
\includegraphics[scale=0.37
]%
{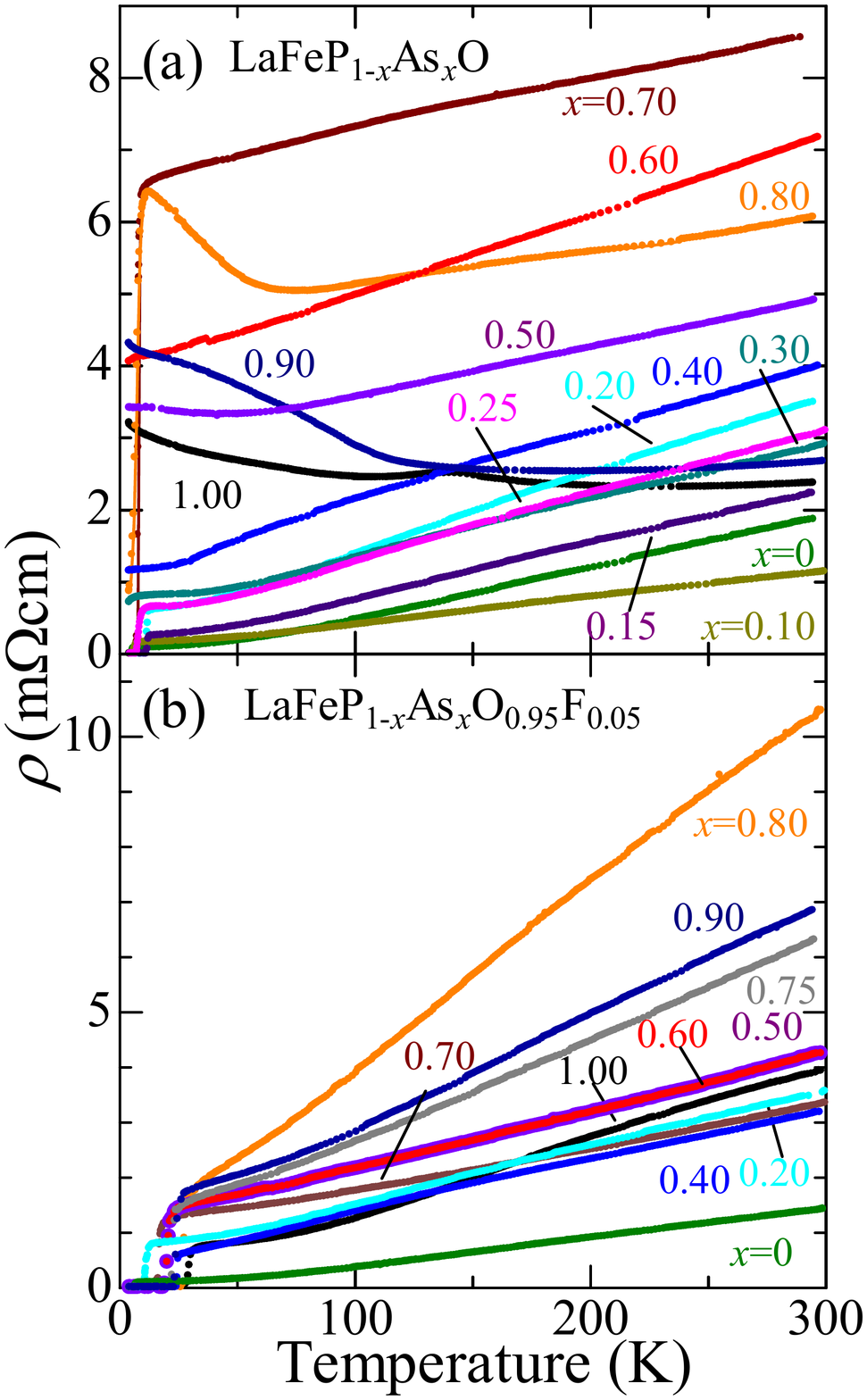}
\caption{(Color online) The temperature dependence of the electrical resistivity of (a) LaFeP$_{1-x}%
$As$_{x}$O and (b) LaFeP$_{1-x}$As$_{x}$O$_{0.95}$F$_{0.05}$ with various values of $x$, respectively.}%
\label{f1}
\end{figure}

Figure \ref{f1} shows the $T$-dependence of resistivity $\rho(T)$ for
the samples of $y$ = 0 and 0.05. The SC transition can be
observed in all the SC samples identified in the magnetic susceptibility measurement. In the samples of $y$ = 0, the anomalous upturn in $\rho(T)$ is
observed for $x$ = 0.8 -- 1 due to the SDW transition, which determines $T_{N}$. The samples of $x$ = 0.3 -- 0.6 seemingly behave as normal metals, but the sample of $x$ = 0.5 shows an upturn at low $T$ ($<$ 50K), which may be related to the recently reported AFM phase \cite{Kitagawa2014}. Actually we observe the broading of NMR spectra at $x$ = 0.3 -- 0.6, which indicates the existence of AFM ordering \cite{Mukuda2}.
We suppose that the behaviors of resistivity obtained in these polycrystalline samples are dominated mainly by the $ab$-plane resistivity $\rho_{ab}$ due to the large ratio of the $c$-axis resistivity $\rho_{c}$ to $\rho_{ab}$. In fact, the large anisotropy ratio $\rho_{c}$/$\rho_{ab}$ ($\sim$20 -- 200) was reported by resistivity measurement of single crystalline LaFeAsO \cite{Jesche2012}.

\begin{figure}[ptb]
\centering
\includegraphics[scale=0.09
]%
{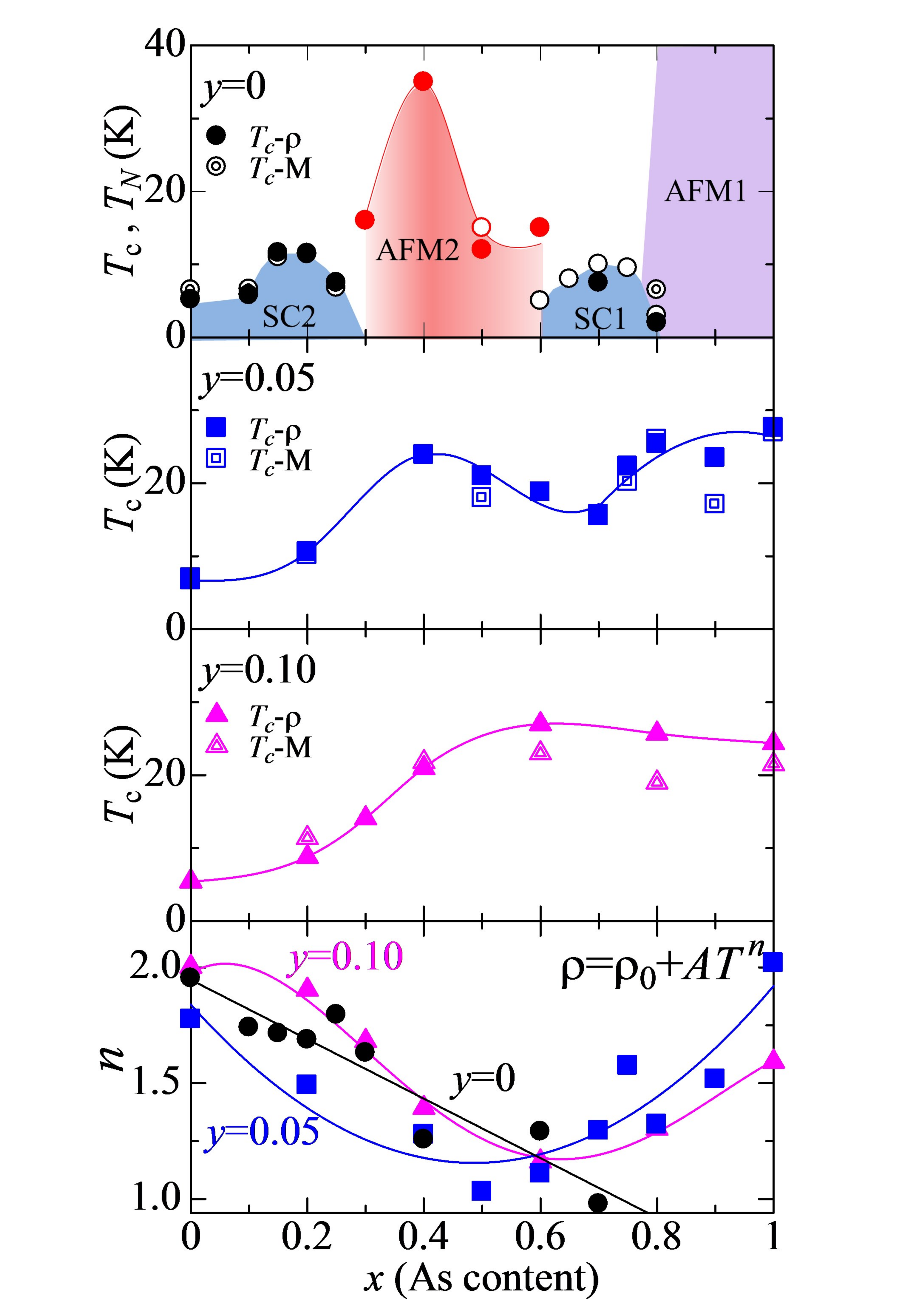}
\caption{(Color online) The As-doping $x$ dependence of (a) critical temperature $T_{c}$,
Neel temperature $T_{N}$ of LaFeP$_{1-x}$As$_{x}$O, (b) $T_{c}$ of LaFeP$_{1-x}$As$_{x}$O$_{0.95}$F$_{0.05}$, (c) $T_{c}$ of LaFeP$_{1-x}$As$_{x}$O$_{0.9}$F$_{0.1}$, and (d) the exponent $n$ in $\rho(T)=\rho_{0}+AT^{n}$ of LaFeP$_{1-x}$As$_{x}$O$_{1-y}$F$_{y}$, respectively. $T_{c}$ is determined by both zero resistivity and the onset of the drop in magnetic susceptibility. The open circles are the data by C. Wang \textit{et al.} \cite{Wang2009} and S. Kitagawa \textit{et al.} \cite{Kitagawa2014}.}%
\label{f2}%
\end{figure}

$T_{c}$ and $T_{N}$ of all the samples for $y$ = 0 are summarized in Fig. \ref{f2}(a). For $y$ =
0, two SC domes and two AFM phases are observed in the phase diagram.
The values of $T_{c}$ at $x$ = 0.6 -- 0.8 (SC1 dome) are consistent with
the previous study \cite{Wang2009}. The other SC dome at $x$ = 0 -- 0.3 (SC2 dome) is first found in the present study. The
maximum $T_{c}$ of SC2 dome is slightly higher ($\sim$12 K) than that of SC1 dome. 
Between the two SC domes, an AFM order (AFM2 phase) is detected via NMR technique \cite{Kitagawa2014,Mukuda2}, with $T_{N}$ ranged from $\sim$15 K to 35 K. Another AFM phase (AFM1 phase) is also observed above $x$ = 0.8 through $\rho(T)$ and NMR \cite{Kitagawa2014}. Here the AFM order is accompanied with a structural phase transition. The values of $T_{N}$, between $\sim$50 K -- 140 K, in AFM1 phase are much higher than that in AFM2 phase.

Fig. \ref{f2}(b) shows the $x$-dependence of $T_{c}$ for $y$ = 0.05. A local minimum of $T_{c}(x)$ is found around $x$ = 0.6, giving a double-peak structure. If we further
increase $y$ to 0.1 \cite{Miyasaka2013}, as illustrated in Fig. \ref{f2}(c), only a single peak is observed at $x$ = 0.6. These results suggest that the two SC domes found at $y$ = 0 merge with each other when $y$ increases.

The exponent $n$ of
$\rho(T)$ is determined by fitting the data with $\rho(T)=\rho_{0}+AT^{n}$
where $\rho_{0}$ is the residual resistivity and $A$ is the slope of $\rho
(T)$. The range of fitting is between $T$ just above $T_{c}$ and $\sim$100 K. As shown in Fig. \ref{f2}(d), for the series of $y$ = 0, $n$ changes gradually from 2 to 1 when $x$ increases to 0.7. Note that the data of $x$ = 0.5, 0.8 -- 1.0 are not suitable for fitting because of the upturn at low $T$. Roughly speaking, the value of $n$ is close to 2 in SC2 dome while that is
$\sim$1 in SC1 dome. It suggests that the behavior of $\rho(T)$ near SC2 dome is described as Fermi liquid (FL)
while that near SC1 dome is non-FL. Since the sample of $x$ = 0.7 is near the boundary of SDW phase, the gradual decrease in $n$ with increasing $x$ suggests the existence of a QCP around $x$ = 0.7. This point of view is
consistent with the previous theoretical prediction \cite{Dai2009}. It is rather surprising that the AFM order observed in NMR does not seriously affect the behaviors in resistivity.

For both series of y=0.05 and 0.1, the value of $n$ 
approaches to 1 around $x$ = 0.6, and to 2 at $x$ = 0 for both series of $y$ = 0.05 and 0.1. It indicates that both systems exhibit non-FL behavior around $x$ = 0.6, while they behave as FL around $x$ = 0.

\begin{figure}[h]%
\centering
\includegraphics[scale=0.34
]%
{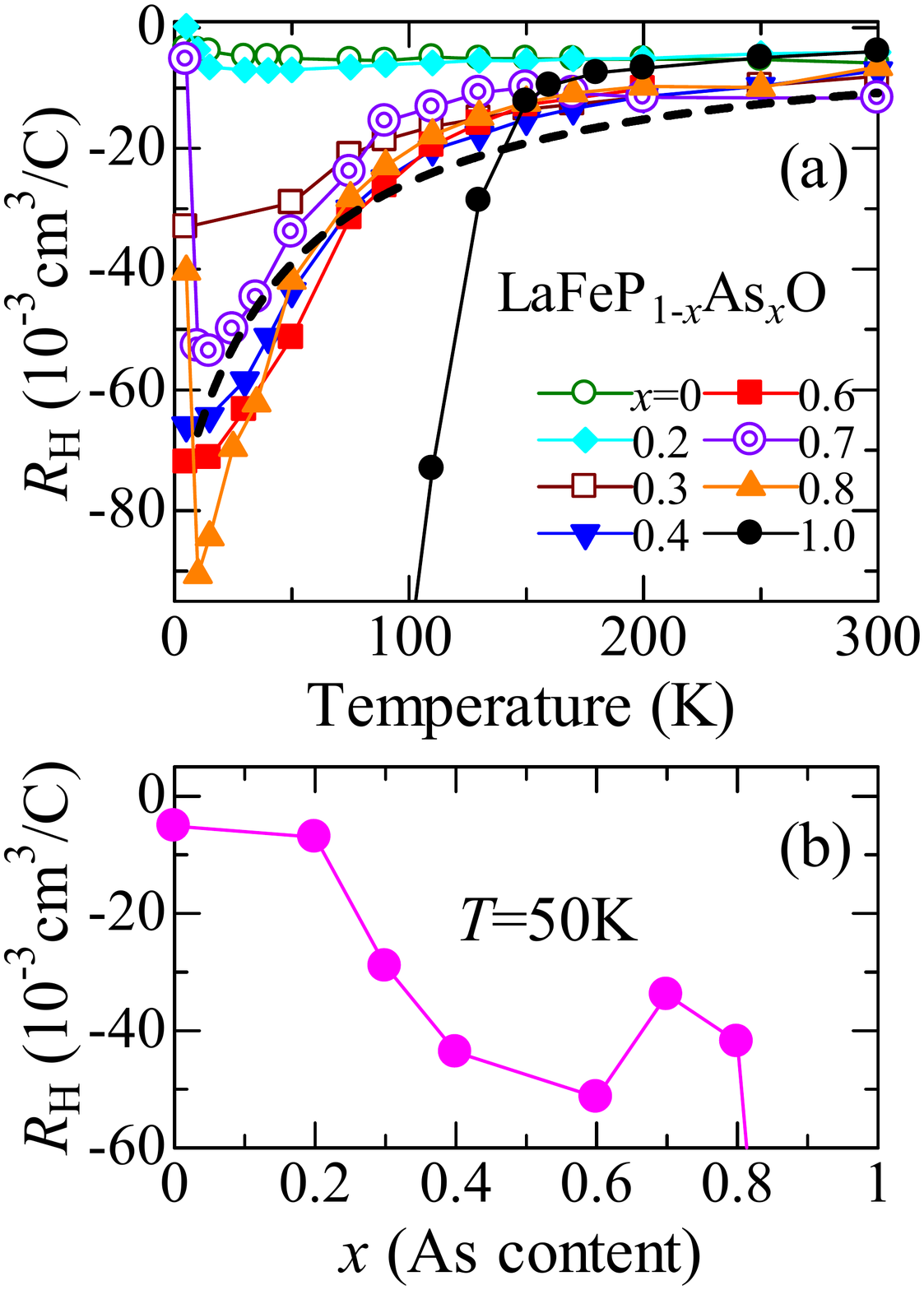}
\caption{(Color online) (a) The temperature dependence of the Hall coefficient $R_{H}$ of
LaFeP$_{1-x}$As$_{x}$O with various $x$. The dashed
curve is the fitting of the data for $x$ = 0.4 as an example, by using the equation $R_{H}$ = -$\alpha_{0}$%
/($T$+$\Theta$). (b) The $x$ dependence of $R_{H}$ at 50 K for
LaFeP$_{1-x}$As$_{x}$O.}%
\label{f3}%
\end{figure}
 
Since Hall coefficient $R_{H}$ is sensitive to the change of electronic states, it has also been measured for LaFeP$_{1-x}$As$_{x}$O. Figure \ref{f3}(a) illustrates $T$
dependence of $R_{H}$ for the series of $y$ = 0. 
For $x$ = 0 and 0.2, $\vert R_{H} \vert$ is very small and almost $T$-independent, which is consistent with the FL picture suggested by the measurements of resistivity. When $x$ exceeds 0.3, $R_{H}$ begins to show a strong $T$-dependence, namely, 
$\vert R_{H} \vert$
is strongly enhanced at low $T$ and suddenly drops at $T_{c}$. Although this low-$T$ enhancement of $\vert R_{H} \vert$ is weakened once at $x$ = 0.7, it increases again at $x \geq$ 0.8.
A large drop of $R_{H}$
observed around 140 K at $x$ = 1 indicates the appearance of
SDW, which is consistent with the $\rho(T)$ data and the previous studies
\cite{Wang2009,Pallecchi2013}. 

To visualize the correspondence to the phase diagram in Fig. \ref{f2}(a), we plot the $x$-dependence of $R_{H}$ at 50 K in Fig. \ref{f3}(b), as a measure of strength of $T$-dependence of $R_H$. Here we can find that SC2 dome (0 $\leq x \leq$ 0.25) shows a weak $T$ dependence of $R_H$, while AFM2 phase (0.3 $< x <$ 0.6) exhibits a strong $T$ dependence. Although the decrease in $R_H$ is slightly recovered at $x$ = 0.7 in SC1 dome, $R_H$ drops again in AFM1 phase above $x$ = 0.8. 

The origin of the strong $T$-dependence of $R_H$ in the intermediate $x$-region may be explained by the following two aspects. The first one is related to the change of FS through P/As
substitution. Band calculation \cite{Kuroki2009} has demonstrated that the main
difference between FS topology of LaFeAsO and LaFePO is the hole pocket
located at ($\pi$,$\pi$,0), named $\gamma$ pocket. Since the $d_{X^{2}-Y^{2}}$
band dominates the FS at ($\pi$,$\pi$,$z$) across $z$ = 0 -- $\pi$ (2D
tube-like FS), the $\gamma$ pocket is observable in LaFeAsO. On the other hand, in LaFePO, the
$d_{Z^{2}}$ band which replaces from $d_{X^{2}-Y^{2}}$ touches the Fermi level only around ($\pi$,$\pi$,$\pi$) but not ($\pi$,$\pi$,0), forming a 3D FS pocket. Therefore, the $\gamma$ pocket in the $k_z$ = 0 plane is absent in LaFePO. If As is substituted by P, the energy of the $d_{X^{2}-Y^{2}}$ and $d_{Z^{2}}$ bands will interchange, and as a result the 
$\gamma$ pocket will shrink in size and finally vanish. We believe that this reconstruction of FS affects the $T$ dependence of $R_{H}$. According to the results in Fig.~\ref{f3}, the interchange of FS topology is supposed to happen between $x$ = 0.3 and 0.8. Following the discussion about the presence of two different kinds of electronic states and their crossover in LaFeP$_{1-x}$As$_{x}$O$_{0.9}$F$_{0.1}$ in our previous study \cite{Miyasaka2013}, we expect that there are two electronic states corresponding to two types of FS topology and the crossover of these two states causes the enhancement of $\vert R_{H} \vert$.

The second aspect is related to the AFM phase between $x$ = 0.3 and 0.6. Since the AFM order may create a charge gap at some part of FS, it may decrease the number of charge carriers and thus enhance $\vert R_{H} \vert$ at low $T$. Another fact is that the $T$ dependent $R_{H}$ of these samples can be roughly fitted with the equation $R_{H}$ =
-$\alpha_{0}$/($T$+$\Theta$), $\alpha_{0}$ and $\Theta$ being some
constants, which is derived from the SF theory \cite{Kontani1999,Nakajima2007}. This relation suggests that the strong $T$ dependence of $\vert R_{H} \vert$
 is related to the presence of the backflow due to strong
electron-electron scattering arising from SF, which is consistent with the observation of low-energy SF above $T_{N}$ in these samples via NMR technique \cite{Kitagawa2014,Mukuda2}. Therefore, the enhancement of $\vert R_{H} \vert$ at $x$ = 0.3 -- 0.6 may be correlated to the 
AFM order in AFM2 phase. A small decrease in $R_H$ at $x$ = 0.7 indicates a recover of FS which gives SC.

The second aspect, however, cannot be adopted in the case of the F-doped system with $y$ = 0.1 \cite{Miyasaka2013}. Although a similar enhancement of $\vert R_{H} \vert$ is observed near $x$ = 0.6, the electronic state is far from the AFM order. Therefore, the anomaly in $R_{H}$ near $x$ = 0.6 for the series of $y$ = 0.1 should be attributed to the band crossover (the first aspect).

\begin{figure}[h]%
\centering
\includegraphics[scale=0.5
]%
{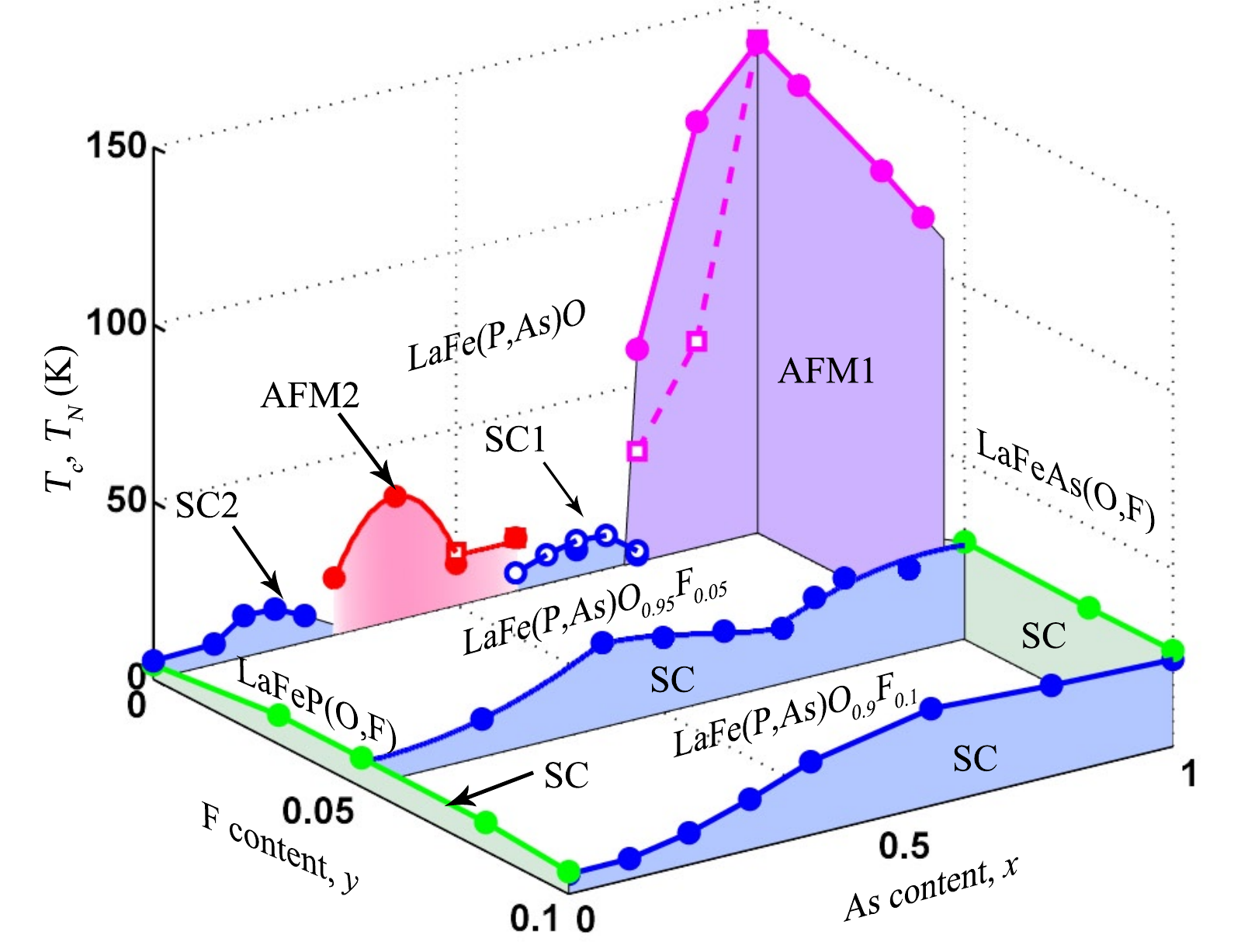}
\caption{(Color online) The phase diagram of LaFeP$_{1-x}$As$_{x}$O$_{1-y}$F$_{y}$ \cite{Suzuki2009,Miyasaka2013,Luetkens2009}. The open dots indicate the data by C. Wang \textit{et al.} \cite{Wang2009} and S. Kitagawa \textit{et al.} \cite{Kitagawa2014}.}%
\label{f4}%
\end{figure}

Finally, we discuss the evolution of the electronic properties in LaFeP$_{1-x}$As$_{x}$O$_{1-y}$F$_{y}$. The phase diagram of this system is
illustrated in Fig. \ref{f4}, which demonstrates the evolution of
two-SC-dome feature. 
The two-dome structure at $y$ = 0 becomes
single-dome structure at $y$ = 0.1. It can be viewed as the
expansion of SC2 dome with increasing $y$ as a result of the suppression of AFM2 phase through F doping. The SF originating from AFM2 phase contributes to the development of SC at $x \sim$ 0.4 in the series of $y$ = 0.05. In the series of $y$ = 0.1, the As-content for the maximum strength of SF and the maximum value of $T_{c}$ is shifted to $x$ = 0.6.
It suggests that SC2 dome further expands and merges with SC1 dome, resulting in a single dome. 

It is clear that the enhancement of $T_{c}$ in SC2 dome (or in the lower $x$ region) is commonly due to the increase of SF suggested by the change of $T$-dependence of $\rho(T)$, namely, the decrease of the exponent $n$ from 2. The enhancement of SF from the FL state has also been observed by NMR experiments \cite{Mukuda2014,Mukuda2}. Therefore, the SC in low-$x$ region is more likely to be induced by SF. In contrast, there is no clear correlation between $T_{c}$ and the exponent $n$ in larger $x$ (As-rich) region.
    
Here we note that the exponent $n$ approaches $\sim$1 around $x$ = 0.6 in all $y$ series (See Figure 2(d)). Although the band crossover is suggested in the region of 0.3 $\leq x \leq$ 0.8 for $y$ = 0 and at $x \sim$ 0.6 for $y$ = 0.1 by the Hall effect measurements, there is no theoretical model that connects  $T$-linear $\rho(T)$ and band crossover. This, together with the SC mechanism in larger $x$ region, is a remaining puzzle.

\section{conclusion}
We have studied the transport properties of polycrystalline
LaFeP$_{1-x}$As$_{x}$O$_{1-y}$F$_{y}$ from $y$ = 0 to 0.1. For $y$ = 0, the new SC dome
(SC2 dome) has been found at $x$ = 0 -- 0.3, constructing a two-dome structure in the phase diagram.
With increasing $y$, SC2 dome expands and merges with SC1 dome. 
Consequently, it results in a double-peak structure of $T_{c}(x)$ in the series of $y$ = 0.05 and a single SC dome in the series of $y$ = 0.1. 
In addition to the AFM phase near LaFeAsO, another AFM phase is observed at $x$ = 0.3 -- 0.6 (AFM2 phase) via NMR in the series of $y$ = 0. Strong temperature dependence of $R_{H}$ observed in AFM2 phase suggests the AFM order opens a charge gap at some parts of FS.
F doping suppresses the AFM order in AFM2 phase and the residual SF induce SC, causing the expansion of SC2 dome. 
The temperature dependence of $\rho(T)$ commonly approaches $T$-linear at $x \sim$ 0.6
in all $y$ series, together with the strong $T$-dependence of $R_{H}$. The evolution of the two SC domes could strongly support the scenario that there exist two different electronic states corresponding to the two FS topologies for the P-rich and the As-rich compositions, respectively. To clarify the SC mechanism for the latter composition regions, further studies are required.

\section*{acknowledgements}
We thank M. Ichimiya for technical support for the EDX measurements. We also thank K. Kuroki for fruitful discussions. This work is supported by JST, TRIP and IRON-SEA.

\end{document}